\documentclass [12pt, letterpaper]{article}
\usepackage{fullpage}
\usepackage{amsmath}
\usepackage{amssymb}
\usepackage{graphicx}
\usepackage{wrapfig}
\usepackage{boxedminipage}
\usepackage{epsfig}
\usepackage{setspace}
\usepackage{subfigure}
\usepackage{float}
\usepackage{hyperref}
\usepackage{soul}
\usepackage{enumerate}

\usepackage{cite}
\newcommand{\be}{\begin{equation}}
\newcommand{\bea}{\begin{eqnarray}}
\newcommand{\eea}{\end{eqnarray}}
\newcommand{\ba}{\begin{align}}
\newcommand{\ea}{\end{align}}
\newcommand{\ee}{\end{equation}}

\begin{document}

\begin{titlepage}
\vspace{10mm}
%\begin{flushright}
 %IPM/P-2015/nnn \\
%FPAUO-1/10\\
%\end{flushright}

\vspace*{20mm}
\begin{center}
{\Large {\bf  Non-local Probes of Entanglement in the Scale-invariant Gravity  
}}

\vspace*{15mm} \vspace*{1mm} {R. Pirmoradian${^1}$ and M. Reza Tanhayi$^{1,2,*}$}

\vspace*{1cm}

{$^1$Department of Physics, Central Tehran Branch, Islamic Azad University  (IAUCTB), P.O. Box 14676-86831,
	Tehran, Iran\\
$^2$School of Physics, Institute for Research in Fundamental Sciences
		(IPM) P.O. Box 19395-5531, Tehran, Iran}

\vspace*{0.5cm}
{E-mail: {\tt $^*$mtanhayi@ipm.ir}}

\vspace*{1cm}
%%\maketitle
\end{center}

\begin{abstract}

In this paper, we study the generic action for the scale-invariant theory of gravity and then  by making use of the holographic methods,  we  compute some specific  holographic measures of entanglement. Precisely, we calculate the entanglement entropy, mutual and tripartite information and show that the mutual information is always positive while the tripartite information becomes negative. This indeed recovers the monogamy property of mutual information in this context.   

\end{abstract}
\tableofcontents
\end{titlepage}

\section{Introduction and Motivation}

It is well known that the gauge principle works well in providing a unified framework for the description of fundamental interactions. For example, by the means of the gauge theory, there is a satisfactory description of the electromagnetic, weak and strong
interactions. But the story is somehow different for gravity since in the gravitational theory the spacetime coordinates appear as the dynamical fields and the gauge principle should be applied to the spacetime
symmetries as well. There is a strong belief that the gauge theory could provide all the interactions including the gravitational interaction. Moreover, large-scale observation indicates that the observed matter in the universe might be described by theories with no fixed time and space scales \cite{Ferreira:2019ywk, Salvio:2014soa}.  In this way, finding scale-invariant theories of gravity seems to be important as long as such theories are in fact gauge theories. \\
Conformal transformation is generated by Poincare and scale (dilatation)
and also special conformal transformations. The Poincare together with the scale transformations form a subgroup named the scale-invariant theory. In a generic situation, considering the scale versus the conformal invariance and also because of the two above mentioned main reasons, finding an acceptable scale-invariant gravity theory has attracted a lot of attention (see for example \cite{Nakayama:2013is}).  In the context of field theory, there is also an important question that in various dimensions whether scale-invariant quantum field theories lead to fully
conformal field theories or not, for example in $d=2$ dimensions see \cite{Iorio:1996ad,Riva:2005gd} and for $d\ne 4$ see \cite{ElShowk:2011gz}. In $d=4$ this subject is interesting and studied extensively. It is claimed that scale-invariant conformal field theory in four dimensions should
also be conformal field theory as well, though there is no comprehensive proof for this claim \cite{Dymarsky:2013pqa,Naseh:2016maw}.\\
% Since its original formulation more than two decade ago, the AdS/CFT correspondence \cite{Maldacena:1997re} has taught us a lot about the quantum nature of gravity. In addition, with the discovery of the Ryu-Takayanagi (RT) formula \cite{Ryu:2006bv} and its generalizations \cite{Dong:2013qoa,Camps:2013zua}, holography and in particular the AdS/CFT correspondence has become and exceedingly important tool in calculating the entanglement entropy (and other entanglement measures) for holographic systems.\\ 
 In this paper, %%we will use a generalization of RT formula for higher derivative theories, due to Dong \cite{Dong:2013qoa}, in order to calculate the entanglement entropy, mutual information, and tripartite information in a scale invariant theory of gravity in four dimensions. In order to arrive at the action of this scale invariant theory, we note that using the procedure of holographic renormalization, the holographic Weyl anomaly in four dimensions is given by 
%\be \langle T^{\mu}_{\mu} \rangle = c_4 \left(R_{\mu\nu}R^{\mu\nu}-\frac{1}{3}R^2\right), \ee
%where $c_4$ is a numerical factor.
we would like to further study different features of scale-invariant gravity in four
dimensions. To proceed, we will first review the way of fixing the corresponding action %as much as we could using certain natural assumptions based on the holographic renormalization \cite{Henningson:1998gx,Anastasiou:2021swo}.  Given the above expression for the Weyl anomaly, one would expect the action 
of the conformal gravity in four dimensions. The theory of conformal gravity, up to dynamically trivial terms, is given by the following action \cite{Riegert:1984zz,Alvarez-Gaume:2015rwa}
\be
I_{\rm CG}=-\frac{\kappa}{16\pi} \int d^4x\sqrt{-g}  \left(R_{\mu\nu}R^{\mu\nu}-\frac{1}{3}R^2\right)+
{\rm total\; derivative\; terms},
\ee
where $\kappa$ is a dimensionless constant. It is well known that the 
total derivatives do not  contribute to the equations of motion, however, such terms play a crucial role when one wants to have a well-defined variational principle. Also, boundary terms contribute to the physical quantities such as the free energy, the entanglement entropy 
and the correlation  function of the holographic stress tensor. Thus in order to explore the scale-invariant theory, it is worth fixing the boundary terms.  In four dimensions, there is a trivial boundary term;  the Gauss-Bonnet term which is a natural total derivative term, after adding this boundary term, the action reads as follows \cite{1}
\be\label{act1}
I_{\rm CG}=-\frac{\kappa}{16\pi} \int d^4x\sqrt{-g}  \left(R_{\mu\nu}R^{\mu\nu}-\frac{1}{3}R^2\right)+\Upsilon \int
d^4x\sqrt{-g}\left(R_{\mu\nu\rho\sigma}R^{\mu\nu\rho\sigma}-4R_{\mu\nu}R^{\mu\nu}+R^2\right),
%\cr
%&&\cr
%&=&-\Upsilon \int
%d^4x\sqrt{-g}\left(R_{\mu\nu\rho\sigma}R^{\mu\nu\rho\sigma}-2R_{\mu\nu}R^{\mu\nu}+\frac{1}{3}
%R^2\right)
\ee
where $\Upsilon$ is an arbitrary constant which we are going to fix it.  To fix the constant $\Upsilon$ we use an asymptotic AdS geometry as a tuner. Indeed since
the AdS geometry has a mathematically well-defined boundary it may be used to fix the boundary
terms. This is what we have learned from the holographic renormalization. An asymptotic AdS geometry may be parameterized as follows\footnote{Note that the first subleasing term in the
	expansion of $g_{ij}$ depends on the theory. For example in Einstein gravity the linear
	term is absent.}
\be
ds^2=\frac{L^2}{r^2}\bigg(dr^2+g_{ij}(x)dx^idx^j\bigg),\;\;\;\;\;\;\;\;{\rm with}\;\;g_{ij}(x)=\delta_{ij}+h_{ij}(x) r+
{\cal O}(r^2).
\ee
Evaluating the action on this solution, usually, leads to divergent terms due to infinite volume
limit (UV divergences)  which could be regularized by adding proper boundary terms. This is,
indeed, what we will do for the model under consideration. Namely, we fix the constant $\Upsilon$
in such a way that the resultant action becomes finite on an asymptotically AdS solution.
Plugging the above solution into action  \eqref{act1} one finds
\be
I_{\rm CG}=-\frac{\kappa}{16\pi} \int d^4x \left(-\frac{12}{r^4}-\frac{a_1}{2 r^2}+
{\cal O}(1)\right)+\Upsilon \int
d^4x\left(\frac{24}{r^4}-\frac{a_1}{ r^2}+
{\cal O}(1)\right),
\ee
where $a_1$ is given in terms of the parameters appearing in the asymptotic
expansion of the metric\cite{Henningson:1998gx}. It is clear that  the on-shell action
is finite if one set $\Upsilon=-\frac{\kappa}{32\pi}$ and the action reads
\be
I_{\rm CG}=-\frac{\kappa}{32\pi}\int d^4x\sqrt{-g}\left(R_{\mu\nu\rho\sigma}R^{\mu\nu\rho\sigma}-2R_{\mu\nu}R^{\mu\nu}+\frac{1}{3}R^2\right),
\ee
which is, indeed, the Weyl squared action. In the next section we will redo the same procedure to write a generic action for the  scale-invariant gravity in four dimensions.

The rest of the paper is organized as follows.  In the next section, we will study the model and find its solutions. In section three, we will further study the mode and use holographic methods to compute holographic entanglement entropy. In section four, we will consider holographic mutual and tripartite information.  Finally, we will briefly discuss our results and some possible future directions in the conclusion part.

\section{ Scale-invariant Gravity in four dimensions}

Let us redo the same procedure as mentioned above for the scale-invariant gravity.
The most general action of a scale-invariant four-dimensional gravity can be written as follows \cite{Alvarez-Gaume:2015rwa}
\bea\label{action-s}
I_{SC}=
-\frac{\kappa}{32\pi}\int d^{4}x\,\sqrt{-g}\,\left(\sigma_0 C_{\mu\nu\alpha\beta}C^{\mu\nu\alpha\beta}
+  R^2+\sigma_1 {\rm GB}_4\right).
\eea
where $C_{\mu\nu\alpha\beta}$ is the Weyl tensor and  $\sigma_0,\sigma_1$ and $\kappa$ are dimensionless constant. ${\rm GB}_4$ stands for the standard Gauss-Bonnet term in four dimensions. Note that we rescale the action
so that the coefficient of $R^2$ term is set to one. As we have already seen the Weyl squared term
is regularized and there is no a divergent term when the action is evaluated on an asymptotically
AdS geometry. Nonetheless, the on-shell action could still be divergent because of the $R^2$ term.
In what follows, we will fix the coefficient of Gauss-Bonnet term $\sigma_1$  such that to remove the divergent
parts of $R^2$ term.\\
To proceed it is important to note that the asymptotic AdS geometry we are
taking should be a solution of the equations of motion. As we have already mentioned this requires
to have certain subleading terms in the asymptotic expansion of the metric.  In the present case,
the corresponding asymptotic behavior is
\be
ds^2=\frac{L^2}{r^2}\bigg(dr^2+g_{ij}(x)dx^idx^j\bigg),\;\;\;\;\;\;\;\;{\rm with}\;\;g_{ij}(x)=\delta_{ij}+h_{ij}(x) r^2+
{\cal O}(r^3).
\ee
For this solution, the on-shell action reads
\bea\label{action-S}
I_{SC}=
-\frac{\kappa}{32\pi}\int d^{4}x\,\left[\sigma_0 {\cal O}(1)+\left(\frac{144}{r^4}
-\frac{6 a_2}{r^2}+{\cal O}(1)\right)+
+ \sigma_1 \left(\frac{24}{r^4}
-\frac{a_2}{r^2}+{\cal O}(1)\right)\right].
\eea
Again $a_2$ is given in terms of the parameters appearing in the asymptotic expansion of the metric \cite{Henningson:1998gx}.  To get a finite action one should set $\sigma_1=-6$. As a result, we will consider the  action of 4D scale-invariant gravity as follows
\bea\label{action-RS}
I_{SC}=
-\frac{\kappa}{32\pi}\int d^{4}x\,\sqrt{-g}\,\left(\sigma_0 C_{\mu\nu\rho\sigma}C^{\mu\nu\rho\sigma}+
R^2-6 {\rm GB}_4\right),
\eea
which is guaranteed to get finite on-shell action for asymptotically AdS geometries. More
explicitly the action may be written in the following form
\bea\label{action-RS1}
I_{SC}=
-\frac{\kappa}{32\pi}\int d^{4}x\,\sqrt{-g}\,\left[(\sigma_0-6)R_{\mu\nu\rho\sigma}^2-2(\sigma_0-12)
R_{\mu\nu}^2+(\frac{\sigma_0}{3}-5) R^2\right].
\eea
To conclude, we note that the most general action of scale-invariant (pure)  gravity in
four dimensions is a  one-family parameter action. This paper aims to study the model
on a generic point of the moduli space of the parameter. In our
notation we recover  conformal gravity at
point $\sigma_0\rightarrow \infty$ while at $\sigma_0=0$ one gets $R^2$ gravity.

The boundary Gauss-Bonnet term does not contribute to the equations of motion and the corresponding  equations derived from the  action \eqref{action-RS1} are as follows
\be\label{EOMT}
\left(\nabla^\sigma\nabla^\rho-\frac{1}{2}R^{\sigma\rho}\right)C_{\mu\sigma\nu\rho}=\frac{1}{2\sigma_0}
\left(R R_{\mu\nu}-g_{\mu\nu}\frac{R^2}{4}-\nabla_\mu\nabla_\nu R+g_{\mu\nu}\Box R\right),
\ee
where $\Box=\nabla^\mu\nabla_\mu$. These equations of motion  admit  several black hole solutions. Indeed, restricting  to Einstein solutions,
the above equations of motion are solved by 
the following black hole solutions (see for example \cite{Kehagias:2015ata})
\be\label{Sol1}
ds^2=\frac{L^2}{r^2}\left(-F(r)\;dt^2+\frac{dr^2}{F(r)}+d\Sigma^2_{2,k}\right),\;\;\;\;\;\;\;
F(r)=\lambda+k  r^2+c_3{r^3},
\ee
where $L$ is the radius of curvature and $k=1,-1,0$ corresponds to $\Sigma_{2,k}= S^2,H_2, R^2$, respectively.  Note that being Einstein solutions they satisfy $R_{\mu\nu}
=\frac{3\lambda}{L^2}g_{\mu\nu}$ with $\lambda=\pm 1,0$. It is worth mentioning that for $\sigma_0>6$ it has Lifshitz solution which is given by 
(see \cite{Lu:2012xu} for Einstein-Weyl gravity)
\be
ds^2=\frac{L^2}{r^2}\left(-\frac{dt^2}{r^{2z}}+dr^2+dx_1^2+dx_2^2\right),\;\;\;\;\;\;\;
z=\frac{\sigma_0-6 +\sqrt{(\sigma_0 -6) (4 \sigma_0 +3)}}{\sigma_0 +3}.
\ee
We note that for $\sigma_0=6$ the action  \eqref{action-RS1} reads
\bea\label{actc}
I_{SI}=
-\frac{3 \kappa}{32\pi}\int d^{4}x\,\sqrt{-g}\,\left[4
R_{\mu\nu}^2- R^2\right],
\eea
which is the curvature squared terms.

%%%%%%%%%%%%%%%%%%%%%%%%%%%%%%%%%%%%%%%%%%%%%%
%%%%%%%%%%%%%%%%%%%%%%%%%%%%%%%%%%%%%%%%%%%%%%%
%%%%%%%%%%%%%%%%%%%%%%%%%%%%%%%%%%%%%%%%%%%%%%%
%%%%%%%%%%%%%%%%%%%%%%%%%%%%%%%%%%%%%%%%%%%%%

\section{ Holographic Entanglement Entropy in the Scale-invariant Theory of Gravity }

As mentioned earlier, the boundary terms contribute to the physical quantities such as the free energy and the entanglement entropy. Thus in order to further explore the role of the Gauss-Bonnet term in 
the  scale-invariant gravity, let us study entanglement entropy. This non-local quantity measures the quantum entanglement between different degrees
of freedom of the system. Similar to other
non-local quantities, \emph{e.g.} Wilson loop and correlation
functions, entanglement entropy can be used to classify the critical points and also the
various quantum phase transitions of the underlying theory.  In quantum field theory,
the entanglement entropy can be defined as follows  \cite{Callan:1994py,Calabrese:2009qy}. In a  $d$ dimensional quantum field theory for a constant time slice, one can define two spatial regions $A$ and its complement $\bar{A}$. The present geometrical division can be translated to the corresponding total Hilbert space as
$\mathcal{H}=\mathcal{H}_A\otimes\mathcal{H}_{\bar{A}}$. For subsystem $A$ the
reduced density matrix is defined by 
tracing the degrees of freedom of ${\bar{A}}$,
as  $\rho_A={\rm Tr}_{\bar{A}}\;\rho$ where $\rho$ is the
total density matrix. The Von
Neumann entropy can give us the entanglement entropy: 
$S_A=-{\rm Tr}_{A}\;\rho_A \log \rho_A$. \\Entanglement entropy becomes infinite due to the infinite correlations
between degrees of freedom near the boundary of the entangling
surface \cite{Bombelli:1986rw, Srednicki:1993im}. Although the entanglement entropy has some
features which are useful to explore the Hilbert space of the theory, but from the quantum field theory point of view computing the entanglement entropy is complicated. Actually, analytic computation of entanglement entropy
is only done in few cases. However, by
virtue of the Anti de Sitter/Conformal Field Theory correspondence
\cite{Maldacena:1997re}, in a seminal work, Ryu and Takayanagi proposed a
simple geometric way to compute the entanglement entropy  \cite{Ryu:2006bv}. In this way, for a $d$ dimensional conformal field theory which
lives on the boundary of an AdS$_{d+1}$ spacetime,  holographic entanglement entropy is given by 
\begin{align}\label{ee}
S_{\rm{EE}}=\frac{\mathcal{A}_{\rm{min}}}{4G_N},
\end{align}
where $G_N$  is the Newton's constant and $\mathcal{A}_{\rm{min}}$ is the minimal surface in the bulk whose border at boundary of the bulk, coincides
with the boundary of the entangling region. Thus the main task is to find the specific minimal surface in the bulk. Since the model under consideration consists of 
 gravity with higher-order terms,  in order to compute the holographic entanglement 
 entropy one should use the generalized Ryu-Takayanagi prescription \cite{{Fursaev:2013fta},{Dong:2013qoa},
 {Camps:2013zua}}.  In particular, for an action containing the most general curvature squared terms   
\be\label{ACT2}
I=-\frac{\kappa}{32\pi}\int d^{4}x\sqrt{-g}\bigg(\lambda_1 R^2+\lambda_2 R_{\mu\nu}
R^{\mu\nu}+\lambda_3 R_{\mu\nu\rho\sigma}R^{\mu\nu\rho\sigma}\bigg),
\ee
the holographic entanglement entropy   
can be obtained by minimizing the following  entropy functional \cite{Fursaev:2013fta}\begin{eqnarray}\label{EE}
S_A\!\!=\!\!\frac{ \kappa}{8}\! \int \!d^2\zeta \;\sqrt{\gamma}\;\bigg[2\lambda_1 R+\lambda_2 \left({ R}_{\mu\nu}n^\mu_i n^\nu_i-\frac{1}{2}\mathcal{K}^i\mathcal{K}_i\right)\!+\!2\lambda_3 \bigg( R_{\mu\nu\rho\sigma}n^\mu_i n^\nu_j n^\rho_i n^\sigma_j-\mathcal{K}^i_{\mu\nu}\mathcal{K}_i^{\mu\nu}\bigg)\!\bigg],
\end{eqnarray}
with $i=1,2$ we denote the two transverse directions to a codimension-two 
hypersurface in the bulk,
$n_i^\mu$  are two unit vectors and ${\cal K}^{(i)}$ stand for the traces of two extrinsic curvature tensors defined by
\be \label{ECUR}
{\cal K}_{\mu\nu}^{(i)}=\pi^\sigma_{\ \mu} \pi^\rho_{\ \nu}\nabla_\rho(n_i)_\sigma,\;\;\;\;\;\;\;
{\rm with}\;\;\;\;  \pi^\sigma_{\ \mu}=\epsilon^\sigma_{\ \mu}+\xi\sum_{i=1,2}(n_i)^\sigma(n_i)_\mu\ ,
\ee
where $\xi=-1$ for space-like and $\xi=1$ for time-like vectors. Moreover,  $\gamma$ is the 
induced metric on the hypersurface with coordinates which are denoted by $\zeta$.  
 
In the scale-invariant theory, let us consider holographic entanglement entropy for a strip as an entangling region. To do so, it is useful 
to parametrize the AdS solution as follows
\be
ds^2=\frac{L^2}{r^2}(-dt^2+dr^2+dx^2+dy^2),
\ee
by which for a constant time slice, the entangling region is given by
\be
t={\rm constant},\;\;\;\;\;\;\;\;\;-\frac{\ell}{2}\leq y\leq \frac{\ell}{2}, \;\;\;\;\;\;\;\;0\leq x\leq H,
\ee
where  $H\gg \ell$ and $H$ play an infrared regulator distance along the entangling surface. The corresponding codimension-two hypersurface in a constant time slice can be parameterized by $y = f\left( r \right)$. The induced metric becomes on the hypersurface becomes (noting that we set $L=1$)
\begin{equation}\label{induced}
d{s^2_{ind}} = \frac{{{1}}}{{{r^2}}}\left[ {\left( {f'{{\left( r \right)}^2} + 1} \right)d{r^2} + d{x}^2 } \right],
\end{equation}
where the $prime$ stands for the derivative with respect to $r.$ Moreover the two-unit vectors normal to the codimension-two hypersurface are 
and the normal vectors are obtained as follows
\begin{equation}\label{unitnormals}
\begin{array}{l}
\Sigma_1\,\,\,:\,\,\,\,t=0\,\,\,\,\,\,\,\,\,\,\,\,\,\,\,\,\,\,\,\,\,\,\,\,\,\,\,\,\,\,{n_1} = \left\{\frac{1}{r },0,0,0\right\}, \\
\Sigma_2\,\,:\,\,\,x-f(r)=0\,\,\,\,\,\,\,\,\,{n_2} = \left\{0,-\frac{ f'}{r \sqrt{f'^2+1}},\,\,\frac{1}{r
	\sqrt{f'^2+1}},\,\,0\right\}. \\
\end{array}
\end{equation}
The corresponding extrinsic curvatures of the hypersurface are given by 
\begin{equation}
{\cal K}_{\mu\nu}^{(1)}=0,\,\,\,\,\,\,\,\,{\cal K}_{\mu\nu}^{(2)}=\left(
\begin{array}{cccc}
0 & 0 & 0 & 0  \\
0 & C_1 & C_1 f' & 0  \\
0 & C_1 f' & C_1  f'^2 & 0 \\
0 & 0 & 0 & C_2  \\

\end{array}
\right)
\end{equation}
where
\begin{eqnarray}
{C_1} = \frac{{2\left( {1 + f{'^2}} \right)f' -2 r f''} }{{2{r ^2}{{(1 + f{'^2})}^{5/2}}}},\,\,\,\,\,\,\,\,\,\,\,\,\,\,\,\,\,\,\,\,\,\,\,\,\,\,\,\,\,\,{C_2} = \frac{{f'}}{{{r ^2}\sqrt {1 + f{'^2}}, }}
\end{eqnarray} Actually, the main task is to minimize the entropy function \eqref{EE} which for a slab entangling region it is given by
\begin{equation}
\begin{array}{l}
S = -\frac{\kappa}{2}\int d r \frac{\sqrt {f'^2 + 1}}{r ^3}\bigg( 1 -40\lambda_1-8\lambda_2-4\lambda_3\\\,\,\,\,\,\,\,\,\,\,\,\,\,\,\,\,\,\,\,\,\,\,\,\,\,\,\,\,- \frac{{{r ^4}}}{{2}}\left[ {\lambda_2{{\left( {2{C_2} + {C_1}\left( {1 + f{'^2}} \right)} \right)}^2} + 4\lambda_3\left( {2{C_2}^2 + {C_1}^2{{\left( {1 + f{'^2}} \right)}^2}} \right)} \right]\bigg),
\end{array}
\end{equation}
Now going back to the scale-invariant case and after replacing the related parameters, the holographic entanglement entropy for the strip is given by the following relation 
\begin{eqnarray}
S=\kappa H \left[-(\sigma_0-6)\left(\frac{1}{\epsilon }-\frac{2 \pi  \Gamma \left(\frac{3}{4}\right)^2}{\Gamma \left(\frac{1}{4}\right)^2}\;\frac{1}{\ell}\right)+(\sigma_0-6)\left(\frac{1}{\epsilon}\right)\right],
\end{eqnarray}
where $\epsilon$ is the UV cut-off of the theory. The first term is the contributions of the dynamical terms, though the last term is that 
of the Gauss-Bonnet term which obviously plays the role of the regulator \cite{Alishahiha:2013dca}. It is also 
interesting to note that in the case of Einstein gravity, the result becomes
\begin{eqnarray}
S_{\rm Ein}=\frac{ H}{2G_N}\left(\frac{1}{\epsilon }-\frac{2 \pi  \Gamma \left(\frac{3}{4}\right)^2}{\Gamma \left(\frac{1}{4}\right)^2}\;\frac{1}{\ell}\right).
\end{eqnarray}
To conclude this section, we have shown that  holographic entanglement entropy for an 
Einstein solution of the action \eqref{action-RS1}  obtained from 
the dynamical part of the action is the same as that of Einstein gravity if we identify $\kappa=\frac{-1}{2(\sigma_0-6)G_N}$ which makes sense as long as $\sigma_0\neq 6$. 

In the following section, we will consider other non-local measurements of entanglement, namely mutual information and tripartite information.

%%%%%%%%%%%%%%%%%%%%%%%%%%%%%%%%%%%%%%%%%%%%%%%%%%%%%%%%%

\section{Holographic mutual and tripartite  Information}

\begin{figure}[H]
	\begin{center}
		\includegraphics[scale=0.62]{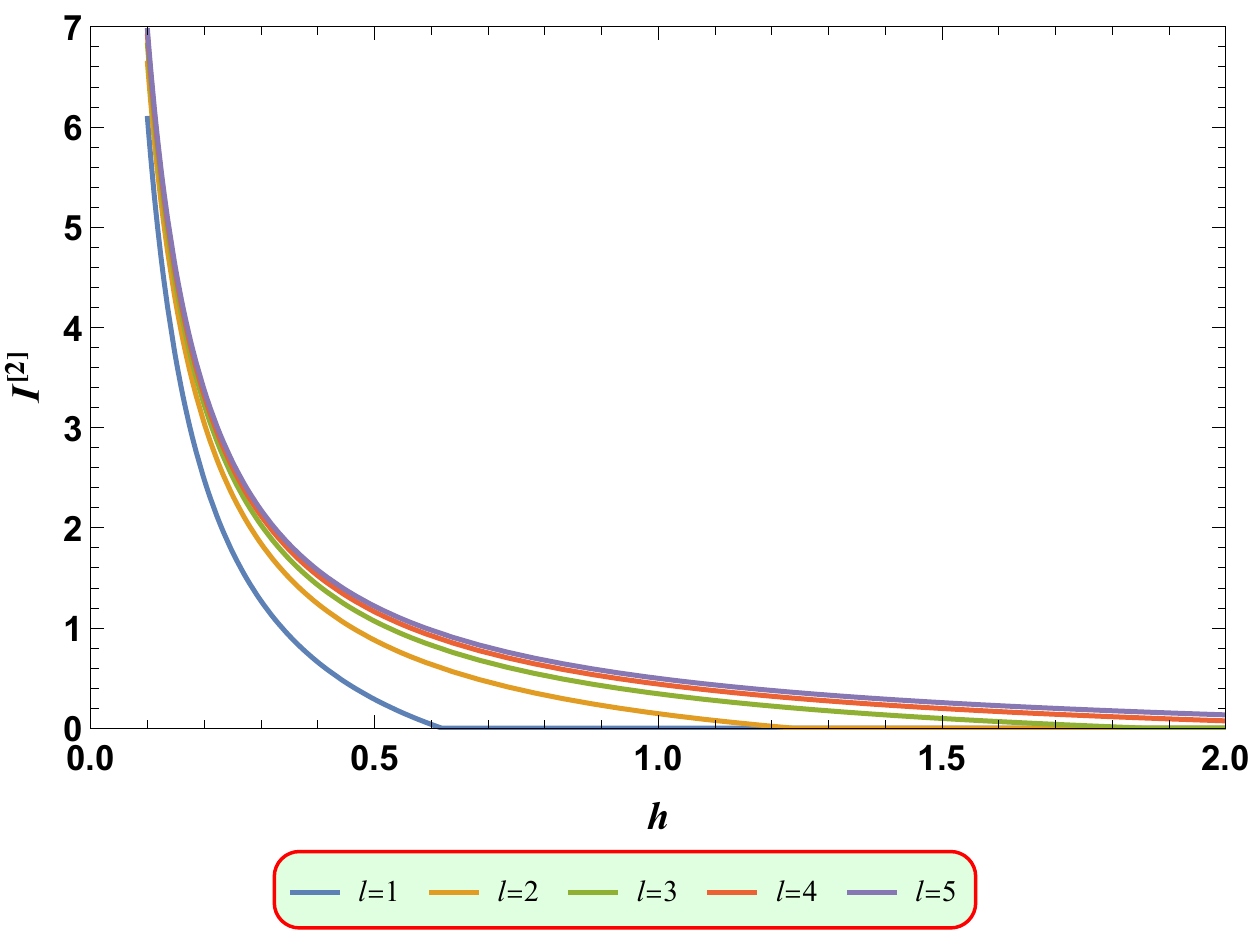}
		\hspace{.5cm}
		\includegraphics[scale=0.62]{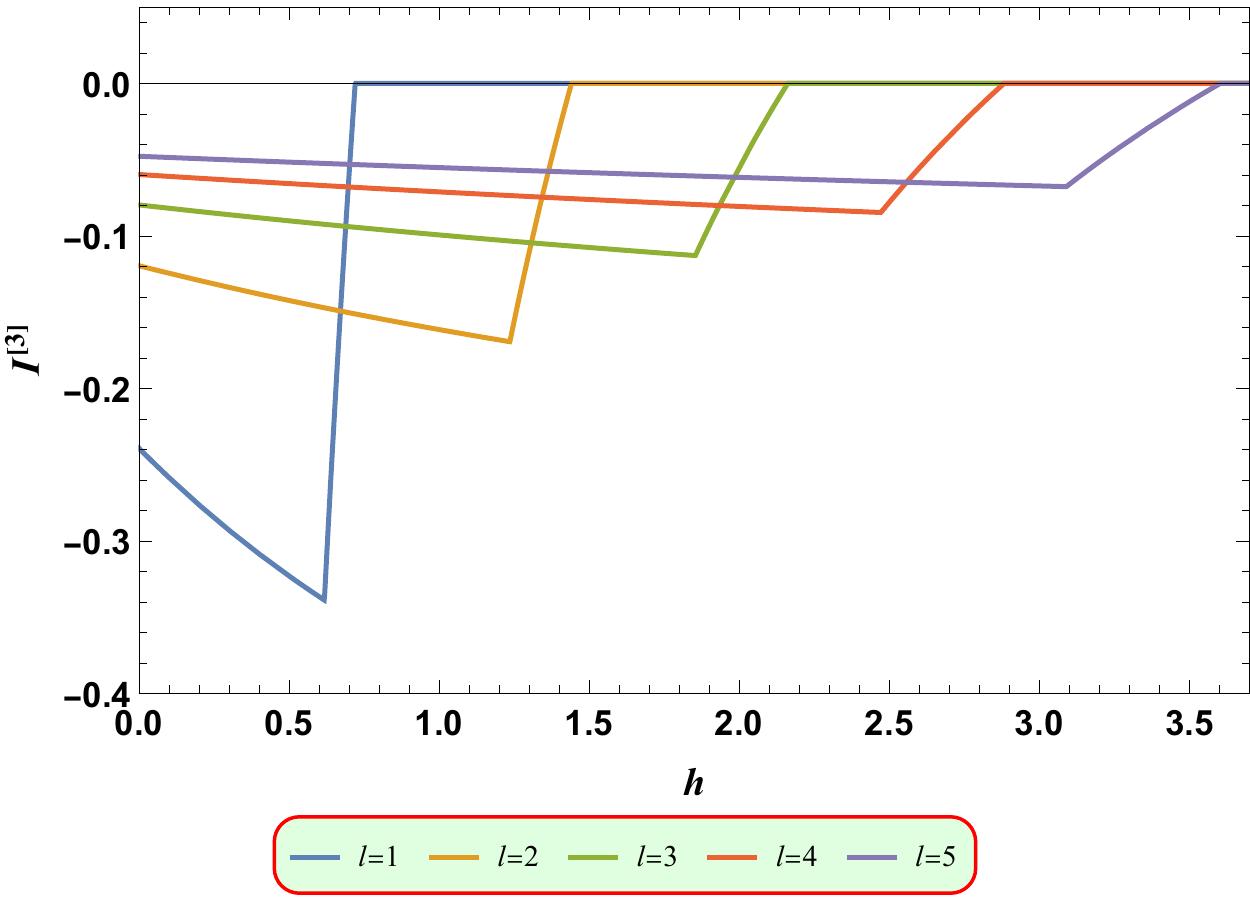}
	\end{center}
	\caption{ Numerical results for holographic mutual information (left plot)  and tripartite information (right plot) as a function of the separation distance: for  ${\ell}=1.1,\cdots, 1.5$. Note in all cases ${\rm{I}}\left({{A},{B}}\right)$ is positive while for three regions the tripartite information remains negative.
	}
	\label{WDWP}
\end{figure}

Entanglement entropy plays a crucial role in
various physical contexts, \emph{e.g.} quantum information theory and black hole physics, however, it suffers some less pleasant features. For example in the generic case, the appearance of the UV cut-off in the expression of
entanglement entropy makes it a non-universal quantity. Thus considering other non-local quantities may help us to improve our
knowledge of the Hilbert space of a quantum system. In the context of quantum information theory, many quantities were
defined to overcome the shortcomings that we encountered using
entanglement entropy. For example, mutual information for a system that has two
disjoint parts is given by\cite{Casini:2008wt}
\begin{align}\label{HMI}
I^{[2]}(\equiv I)(A,B)=S(A)+S(B)-S(A\cup B),
\end{align}
 where  $S(A_i)$'s are the entanglement entropies and $S(A\cup B)$ is the entanglement entropy for the union
of the two entangling regions. This is a finite quantity and
quantifies the amount of information which is
shared between two subsystems. intuitively, for three subregions, the tripartite information is defined by
\begin{align}\label{3par}
{I^{[3]}}\left( {{A},{B},{C}} \right) &= S\left( {{A}} \right) + S\left( {{B}} \right) + S\left( {{C}} \right) - S\left( {{A} \cup {B}} \right) - S\left( {{A} \cup {C}} \right) \notag\\
&- S\left( {{B} \cup {C}} \right) + S\left( {{A} \cup {B} \cup {C}} \right).
\end{align}
In the above relations, $S\left( {{A_i} \cup {A_j}} \right)$ is the entanglement entropy for the union of the entangling regions.  In this section, we use holographic methods to compute holographic mutual and tripartite information.  It is worth mentioning that in writing the tripartite information, the union parts play an important role. To explore this point let us write the tripartite information in terms of the mutual information as follows
\begin{align}\label{3par1}
{I^{[3]}}({A},{B},{C}) &= I({A} , {B}) + I({A}, {C}) - I({A} ,{B} \cup {C}),
\end{align}
which helps us to investigate the sign of mutual information. 
Now the aim is to compute the mutual information. For two strips with the same length of $\ell$ separated by distance $h$, after making use of the corresponding holographic entanglement entropy, the holographic mutual information is given by  
\begin{equation}\label{mutualcon}
I\left( {{A},{B}} \right) =\frac{4\pi H}{G_N}\frac{\Gamma(\frac{3}{4})^2}{\Gamma(\frac{1}{4})^2}
\Big(\frac{1}{2\ell+h}+\frac{1}{h}-\frac{2}{\ell}\Big).
\end{equation}
\begin{figure}[H]
	\centering
	\includegraphics[width=10cm]{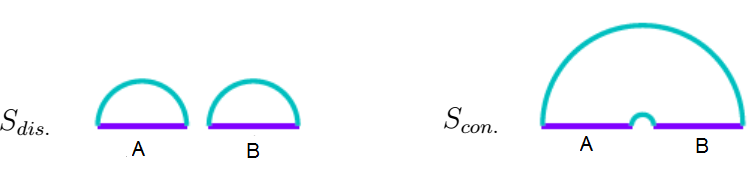}\,\,\,\,\,\,\,\,\,\,\,\,%\includegraphics[width=6cm]{Schematic-Right}
	\caption{Schematic representation of two different configurations for computing the entanglement entropy of union of regions.}
\end{figure}
\begin{figure}[H]
\centering 	\includegraphics[width=15cm]{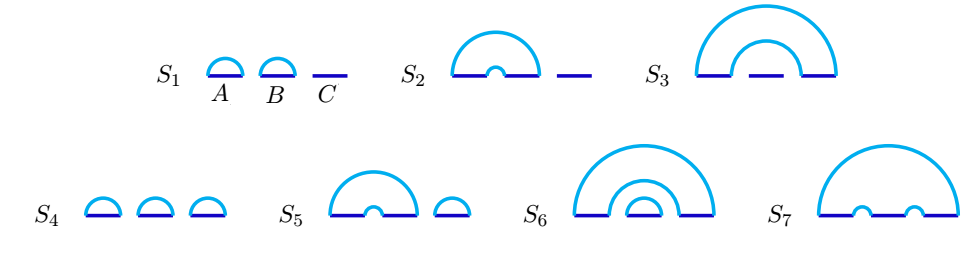}

\caption{ Schematic representation of competing configurations for three subregions. In the computation of $S(A\cup B)$ and $S\left( {{A}\cup{B}\cup{C}} \right)$ there are some configurations, where the minimal one must be used. }
\label{imutual1}
\end{figure} 
On the other hand, for three entangling regions with the same length $\ell$ separated by distance $h$, the main point is finding the union part of entanglement entropy. According to the holographic principle, a minimal configuration in the bulk space is needed. In Fig.\ref{imutual1}, we have plotted all possible diagrams of the union of three regions and $S(A_i\cup A_j)$ and $S\left( {{A}\cup{B}\cup{C}} \right)$  are given by the minimum among the possible diagrams. For two and three strips as the entangling regions with the same length, $\ell$ one can write \[\begin{array}{l}
S\left( {{A_i} \cup {A_j}} \right):\left\{ \begin{array}{l}
2S\left( \ell  \right) \equiv {S_1} \\
S\left( {2\ell  + h} \right) + S\left( h \right) \equiv {S_2} \\
S\left( {3\ell  + 2h} \right) + S\left( {\ell  + 2h} \right) \equiv {S_3} \\
\end{array} \right.
\end{array}\] 
and for three entangling regions one has 
\[\begin{array}{l}
S\left( {{A} \cup{B} \cup{C}} \right):\left\{ \begin{array}{l}
3S\left( \ell  \right) \equiv {S_4} \\
S\left( {3\ell  + 2h} \right) + S\left( {\ell  + 2h} \right) + S\left( \ell  \right) \equiv {S_5} \\
S\left( {2\ell  + h} \right) + S\left( \ell  \right) + S\left( h \right) \equiv {S_6} \\
S\left( {3\ell  + 2h} \right) + 2S\left( h \right) \equiv {S_7} \\
\end{array} \right. \\
\end{array}\]
From these possible configurations, and after making use of the minimum expression in each case, the holographic tripartite information is obtained as follows
\begin{equation}
{I^{\left[ 3 \right]}}\left( {A,B,C} \right) = 3S\left( \ell  \right) - 2\min \left\{ {{S_1},{S_2}} \right\} - \min \left\{ {{S_1},{S_3}} \right\} + \min \left\{ {{S_4},{S_5},{S_6},{S_7}} \right\},
\end{equation}
noting that by $\min \left\{ {{S_1},{S_2}} \right\}$ we mean the minimum configuration between $S_1$ and $S_2$.  \\
In the theory that we are dealing with, in figure \ref{WDWP}, we have plotted numerically the mutual and tripartite information for some certain value of $\ell$ and $h$. The numerical results indicate that for the parameters that we have used, the mutual information is always positive, on the other hand, the tripartite information remains negative. 

%%%%%%%%%%%%%%%%%%%%%%%%%%%%%%%%%%%%%%%%%%%%%%%%%%%%%%%%%%%%%%%%%%%%%

\section{Concluding Remarks}

In this paper, we studied different features of holographic entanglement measures for the scale-invariant gravity in four dimensions. The action of the scale-invariant theory can indeed be parameterized by one family of free parameters say as $\sigma_0$. And for $\sigma_0\rightarrow \infty$, the theory reduces to the four-dimensional conformal gravity, while at $\sigma_0=0$ it reduces to pure $R^2$ gravity. In this theory, we studied the holographic entanglement entropy and  mutual and tripartite information.  Actually, in the scale-invariant theory of gravity, for entangling regions $A,B,C$ in the boundary field theory, the tripartite information obeys the following inequality
\begin{equation}
I^{[3]}(A,B,C) \le 0.
\end{equation}
The tripartite information is always negative. From the definition of the tripartite information, one can write the tripartite information in terms of the mutual information as the equation \eqref{3par1}. An immediate result is 
\begin{align}
I({A} , {B}) + I({A}, {C}) \le
I({A} , {B} \cup {C}),
\end{align} that means the sign of tripartite information constraints the mutual information. This is equivalent to the monotonicity of mutual information implies certain bounds when applied to the long-distance expansion of the mutual information.\\
In the context of quantum information theory, for any measure of the information say as $F(A)$, the inequality of form $F(A, B)+F(A, C) \le F(A, B\cup C)$
is known as monogamy relation indicating that the holographic mutual information is monogamous. This feature is characteristic of measures of quantum entanglement. In the context of quantum information theory, the monogamy property is related to the security of quantum cryptography, noting that quantum entanglement, unlike the classical correlation, is not a shareable resource. In other words, entangled correlations between $A$ and $B$ cannot be shared with a third system $C$ without spoiling the original entanglement \cite{Hayden:2011ag, Mirabi:2016elb}. We showed that scale-invariant gravity also respects the monogamy feature of the information. 

As future work, we will study the other solutions of the scale-invariant gravity in four dimensions. For example, the theory that we have considered,  
for $\sigma_0=6$, exhibits a new solution. In this critical point, all  entanglement entropies that we have
computed  are identically zero. This indicates that at this critical point the model may have a   logarithmic vacuum solution. It is  worth mentioning that adding  the log term to an AdS solution may be identified to a deformation of the dual conformal field theory by an irrelevant operator, this is what we have learned form the gauge/gravity  duality.  Therefore, adding the log term might destroy the conformal symmetry of the model at UV, therefore studying the way of  applying the duality of gauge/gravity becomes important.  Nevertheless, following \cite{Skenderis:2009nt}, one may assume that the deformation  is sufficiently small and this term may be treated perturbatively. We leave the further investigation to other works.

{\section*{Acknowledgments}
	
We would like to thank M. Alishahiha for bringing out attention on his work \cite{1} about the scale-invariant theory.  So special thanks to him on behalf of all his supporting and  generosity and also for his useful comments and hints. We would like to thank R. Vazirian for her helpful comments.  R.P. would also like to thank A. Naseh and B. Taghavi for useful conversations and comments.

\section*{Appendix: Some useful mathematical relations}

Here in this appendix, we present some useful relations that we have used in this paper. Let us choose a five-dimensional black hole solution with coordinate $t, r,x,y,z$ as follows 
\begin{equation*}
\left(
\begin{array}{ccccc}
-\frac{f(r)}{r^2} & 0 & 0 & 0 & 0 \\
0 & \frac{1}{r^2 f(r)} & 0 & 0 & 0 \\
0 & 0 & \frac{1}{r^2} & 0 & 0 \\
0 & 0 & 0 & \frac{1}{r^2} & 0 \\
0 & 0 & 0 & 0 & \frac{1}{r^2} \\
\end{array}
\right)
\end{equation*}
The determinant of the induced metric reads as 
\begin{equation*}
\frac{1}{r^6 f(r)}+\frac{x'(r)^2}{r^6}
\end{equation*}
Therefore two normal vectors are obtained as 
\begin{equation*}
\left\{\frac{1}{\sqrt{\frac{r^2}{f(r)}}},0,0,0,0\right\},\,\,\,\,\left\{0,-\frac{x'(r)}{\sqrt{r^2 f(r) x'(r)^2+r^2}},\frac{1}{\sqrt{r^2 f(r)
		x'(r)^2+r^2}},0,0\right\}
\end{equation*}
The non-zero component of the extrinsic curvature ten reads as
$${\cal K}_{11}=\frac{-r f'(r) x'(r)+2 f(r)^2 x'(r)^3+2 f(r) \left(x'(r)-r x''(r)\right)}{2 r^2 f(r)
	\left(f(r) x'(r)^2+1\right)^{5/2}}$$
$${\cal K}_{12}=\frac{x'(r) \left(-r f'(r) x'(r)+2 f(r)^2 x'(r)^3+2 f(r) \left(x'(r)-r
	x''(r)\right)\right)}{2 r^2 \left(f(r) x'(r)^2+1\right)^{5/2}}$$
$${\cal K}_{21}=\frac{x'(r) \left(-r f'(r) x'(r)+2 f(r)^2 x'(r)^3+2 f(r) \left(x'(r)-r
	x''(r)\right)\right)}{2 r^2 \left(f(r) x'(r)^2+1\right)^{5/2}}$$
$${\cal K}_{22}=\frac{f(r) x'(r)^2 \left(-r f'(r) x'(r)+2 f(r)^2 x'(r)^3+2 f(r) \left(x'(r)-r
	x''(r)\right)\right)}{2 r^2 \left(f(r) x'(r)^2+1\right)^{5/2}}$$
$${\cal K}_{33}=\frac{f(r) x'(r)}{r^2 \sqrt{f(r) x'(r)^2+1}}$$
\begin{equation*}
{\cal K}_{44}=\frac{f(r) x'(r)}{r^2 \sqrt{f(r) x'(r)^2+1}}
\end{equation*}
and also one finds
\begin{equation*}
{R_{\mu \nu }}n_i^\mu n_i^\nu =\frac{r f'(r)-4 f(r)}{f(r) x'(r)^2+1}-\frac{f(r) x'(r)^2 \left(r^2 f''(r)-5 r f'(r)+8
	f(r)\right)}{2 \left(f(r) x'(r)^2+1\right)}+\frac{1}{2} \left(-r \left(r f''(r)-5
f'(r)\right)-8 f(r)\right)
\end{equation*}

\begin{equation*}
{R_{\mu \nu \alpha \beta }}n_i^\mu n_i^\alpha n_j^\nu n_j^\beta=-2 \left(\frac{f(r) x'(r)^2 \left(r^2 f''(r)-2 r f'(r)+2 f(r)\right)}{2 \left(f(r)
	x'(r)^2+1\right)}-\frac{r f'(r)-2 f(r)}{2 \left(f(r) x'(r)^2+1\right)}\right) 
\end{equation*}

%For $$ \lambda_1=\lambda_3=\frac{\lambda}{2}\hspace{2mm} and\hspace{2mm}  \lambda_2=-2 \lambda$$ We will get \begin{equation*} GB=\frac{2 \lambda  r f'(r)-2 \lambda  f(r) \left(f(r) x'(r) \left(3 x'(r)+2 r x''(r)\right)+3\right)}{\left(f(r) x'(r)^2+1\right)^2} \end{equation*}

%\begin{equation*} entropyintegrand=\sqrt{\frac{f(r) x'(r)^2+1}{r^6 f(r)}} \left(\frac{2 \lambda  r f'(r)-2 \lambda  f(r) \left(f(r) x'(r) \left(3 x'(r)+2 r x''(r)\right)+3\right)}{\left(f(r) x'(r)^2+1\right)^2}+1\right) \end{equation*} 
After doing some straightforward algebra for a strip entangling region  the integrand function becomes
\begin{equation*}
{\rm strip\,\, entegrand }=-\frac{x'(r) \left(f(r) \left(x'(r)^2-2 \lambda \right)+1\right)}{r^3 f(r) \left(\frac{1}{f(r)}+x'(r)^2\right)^{3/2}},
\end{equation*}
where we have used $ \lambda_1=\lambda_3=\frac{\lambda}{2}$  and $ \lambda_2=-2 \lambda$.


\begin{thebibliography}{}
	
\bibitem{Ferreira:2019ywk}
P.~G.~Ferreira and O.~J.~Tattersall,
%``Scale Invariant Gravity and Black Hole Ringdown,''
Phys. Rev. D \textbf{101} (2020) no.2, 024011
doi:10.1103/PhysRevD.101.024011
[arXiv:1910.04480 [gr-qc]].
%2 citations counted in INSPIRE as of 20 Mar 2021

\bibitem{Salvio:2014soa}
A.~Salvio and A.~Strumia,
%``Agravity,''
JHEP \textbf{06} (2014), 080
doi:10.1007/JHEP06(2014)080
[arXiv:1403.4226 [hep-ph]].
%237 citations counted in INSPIRE as of 06 Apr 2021

	
	
\bibitem{Nakayama:2013is}
Y.~Nakayama,
%``Scale invariance vs conformal invariance,''
Phys. Rept. \textbf{569} (2015), 1-93
doi:10.1016/j.physrep.2014.12.003
[arXiv:1302.0884 [hep-th]].
%227 citations counted in INSPIRE as of 02 Apr 2021




\bibitem{Iorio:1996ad}
A.~Iorio, L.~O'Raifeartaigh, I.~Sachs and C.~Wiesendanger,
%``Weyl gauging and conformal invariance,''
Nucl. Phys. B \textbf{495} (1997), 433-450
doi:10.1016/S0550-3213(97)00190-9
[arXiv:hep-th/9607110 [hep-th]].
%130 citations counted in INSPIRE as of 26 Mar 2021
 

 \bibitem{Riva:2005gd}
 V.~Riva and J.~L.~Cardy,
 %``Scale and conformal invariance in field theory: A Physical counterexample,''
 Phys. Lett. B \textbf{622} (2005), 339-342
 doi:10.1016/j.physletb.2005.07.010
 [arXiv:hep-th/0504197 [hep-th]].
 %70 citations counted in INSPIRE as of 20 Mar 2021


\bibitem{ElShowk:2011gz}
S.~El-Showk, Y.~Nakayama and S.~Rychkov,
%``What Maxwell Theory in D\ensuremath{<}\ensuremath{>}4 teaches us about scale and conformal invariance,''
Nucl. Phys. B \textbf{848} (2011), 578-593
doi:10.1016/j.nuclphysb.2011.03.008
[arXiv:1101.5385 [hep-th]].
%97 citations counted in INSPIRE as of 20 Mar 2021


\bibitem{Dymarsky:2013pqa}
A.~Dymarsky, Z.~Komargodski, A.~Schwimmer and S.~Theisen,
%``On Scale and Conformal Invariance in Four Dimensions,''
JHEP \textbf{10} (2015), 171
doi:10.1007/JHEP10(2015)171
[arXiv:1309.2921 [hep-th]].
%101 citations counted in INSPIRE as of 20 Mar 2021


\bibitem{Naseh:2016maw}
A.~Naseh,
%``Scale versus conformal invariance from entanglement entropy,''
Phys. Rev. D \textbf{94} (2016) no.12, 125015
doi:10.1103/PhysRevD.94.125015
[arXiv:1607.07899 [hep-th]].
%11 citations counted in INSPIRE as of 20 Mar 2021

\bibitem{Riegert:1984zz}
R.~J.~Riegert,
%``Birkhoff's Theorem in Conformal Gravity,''
Phys. Rev. Lett. \textbf{53} (1984), 315-318
doi:10.1103/PhysRevLett.53.315
%136 citations counted in INSPIRE as of 07 Apr 2021


\bibitem{Alvarez-Gaume:2015rwa}
L.~Alvarez-Gaume, A.~Kehagias, C.~Kounnas, D.~L\"ust and A.~Riotto,
%``Aspects of Quadratic Gravity,''
Fortsch. Phys. \textbf{64} (2016) no.2-3, 176-189
doi:10.1002/prop.201500100
[arXiv:1505.07657 [hep-th]].
%124 citations counted in INSPIRE as of 06 Apr 2021


\bibitem{1} M. Alishahiha, ``On 4D Scale Invariant Gravity,'' unpublished.


\bibitem{Henningson:1998gx}
M.~Henningson and K.~Skenderis,
%``The Holographic Weyl anomaly,''
JHEP \textbf{07} (1998), 023
doi:10.1088/1126-6708/1998/07/023
[arXiv:hep-th/9806087 [hep-th]].
%1439 citations counted in INSPIRE as of 01 Apr 2021


%\bibitem{Ghodsi:2014hua}
%A.~Ghodsi, B.~Khavari and A.~Naseh,
%``Holographic Two-Point Functions in Conformal Gravity,''
%JHEP \textbf{01} (2015), 137
%doi:10.1007/JHEP01(2015)137
%[arXiv:1411.3158 [hep-th]].
%12 citations counted in INSPIRE as of 20 Mar 202


\bibitem{Kehagias:2015ata}
A.~Kehagias, C.~Kounnas, D.~L\"ust and A.~Riotto,
%``Black hole solutions in $R^{2}$ gravity,''
JHEP \textbf{05} (2015), 143
doi:10.1007/JHEP05(2015)143
[arXiv:1502.04192 [hep-th]].
%51 citations counted in INSPIRE as of 29 Mar 2021

\bibitem{Lu:2012xu}
H.~Lu, Y.~Pang, C.~N.~Pope and J.~F.~Vazquez-Poritz,
%``AdS and Lifshitz Black Holes in Conformal and Einstein-Weyl Gravities,''
Phys. Rev. D \textbf{86} (2012), 044011
doi:10.1103/PhysRevD.86.044011
[arXiv:1204.1062 [hep-th]].
%150 citations counted in INSPIRE as of 20 Mar 2021


\bibitem{Callan:1994py}
C.~G.~Callan, Jr. and F.~Wilczek,
%``On geometric entropy,''
Phys. Lett. B \textbf{333} (1994), 55-61
doi:10.1016/0370-2693(94)91007-3
[arXiv:hep-th/9401072 [hep-th]].
%607 citations counted in INSPIRE as of 06 Apr 2021


\bibitem{Calabrese:2009qy}
P.~Calabrese and J.~Cardy,
%``Entanglement entropy and conformal field theory,''
J. Phys. A \textbf{42} (2009), 504005
doi:10.1088/1751-8113/42/50/504005
[arXiv:0905.4013 [cond-mat.stat-mech]].
%708 citations counted in INSPIRE as of 06 Apr 2021


\bibitem{Bombelli:1986rw}
L.~Bombelli, R.~K.~Koul, J.~Lee and R.~D.~Sorkin,
%``A Quantum Source of Entropy for Black Holes,''
Phys. Rev. D \textbf{34} (1986), 373-383
doi:10.1103/PhysRevD.34.373
%1063 citations counted in INSPIRE as of 07 Apr 2021


\bibitem{Srednicki:1993im}
M.~Srednicki,
%``Entropy and area,''
Phys. Rev. Lett. \textbf{71} (1993), 666-669
doi:10.1103/PhysRevLett.71.666
[arXiv:hep-th/9303048 [hep-th]].
%1213 citations counted in INSPIRE as of 06 Apr 2021


\bibitem{Maldacena:1997re}
J.~M.~Maldacena,
%``The Large N limit of superconformal field theories and supergravity,''
Adv. Theor. Math. Phys. \textbf{2} (1998), 231-252
doi:10.1023/A:1026654312961
[arXiv:hep-th/9711200 [hep-th]].
%16515 citations counted in INSPIRE as of 07 Apr 2021


\bibitem{Ryu:2006bv}
S.~Ryu and T.~Takayanagi,
%``Holographic derivation of entanglement entropy from AdS/CFT,''
Phys. Rev. Lett. \textbf{96} (2006), 181602
doi:10.1103/PhysRevLett.96.181602
[arXiv:hep-th/0603001 [hep-th]].
%2610 citations counted in INSPIRE as of 06 Apr 2021


\bibitem{Dong:2013qoa}
X.~Dong,
%``Holographic Entanglement Entropy for General Higher Derivative Gravity,''
JHEP \textbf{01} (2014), 044
doi:10.1007/JHEP01(2014)044
[arXiv:1310.5713 [hep-th]].
%293 citations counted in INSPIRE as of 06 Apr 2021


\bibitem{Camps:2013zua}
J.~Camps,
%``Generalized entropy and higher derivative Gravity,''
JHEP \textbf{03} (2014), 070
doi:10.1007/JHEP03(2014)070
[arXiv:1310.6659 [hep-th]].
%188 citations counted in INSPIRE as of 29 Mar 2021


\bibitem{Fursaev:2013fta}
D.~V.~Fursaev, A.~Patrushev and S.~N.~Solodukhin,
%``Distributional Geometry of Squashed Cones,''
Phys. Rev. D \textbf{88} (2013) no.4, 044054
doi:10.1103/PhysRevD.88.044054
[arXiv:1306.4000 [hep-th]].
%142 citations counted in INSPIRE as of 06 Apr 2021


\bibitem{Alishahiha:2013dca}
M.~Alishahiha, A.~F.~Astaneh and M.~R.~Mohammadi Mozaffar,
%``Holographic Entanglement Entropy for 4D Conformal Gravity,''
JHEP \textbf{02} (2014), 008
doi:10.1007/JHEP02(2014)008
[arXiv:1311.4329 [hep-th]].
%13 citations counted in INSPIRE as of 20 Mar 2021


\bibitem{Casini:2008wt}
H.~Casini and M.~Huerta,
%``Remarks on the entanglement entropy for disconnected regions,''
JHEP \textbf{03} (2009), 048
doi:10.1088/1126-6708/2009/03/048
[arXiv:0812.1773 [hep-th]].
%130 citations counted in INSPIRE as of 20 Mar 2021


\bibitem{Hayden:2011ag}
P.~Hayden, M.~Headrick and A.~Maloney,
%``Holographic Mutual Information is Monogamous,''
Phys. Rev. D \textbf{87} (2013) no.4, 046003
doi:10.1103/PhysRevD.87.046003
[arXiv:1107.2940 [hep-th]].
%213 citations counted in INSPIRE as of 20 Mar 2021

%\cite{Mirabi:2016elb}
\bibitem{Mirabi:2016elb}
S.~Mirabi, M.~R.~Tanhayi and R.~Vazirian,
%``On the Monogamy of Holographic $n$-partite Information,''
Phys. Rev. D \textbf{93}, no.10, 104049 (2016)
doi:10.1103/PhysRevD.93.104049
[arXiv:1603.00184 [hep-th]].

\bibitem{Skenderis:2009nt}
K.~Skenderis, M.~Taylor and B.~C.~van Rees,
%``Topologically Massive Gravity and the AdS/CFT Correspondence,''
JHEP \textbf{09} (2009), 045
doi:10.1088/1126-6708/2009/09/045
[arXiv:0906.4926 [hep-th]].
%127 citations counted in INSPIRE as of 20 Mar 2021




\end{thebibliography}
\end{document}